\documentstyle[preprint,aps,psfig]{revtex}
\tighten

\begin{document}
\draft

\preprint{\small TPI-MINN-99/33, $\;$
                              UMN-TH-1807-99}

\newcommand{\nc}{\newcommand}
\nc{\al}{\alpha}
\nc{\ga}{\gamma}
\nc{\de}{\delta}
\nc{\ep}{\epsilon}
\nc{\ze}{\zeta}
\nc{\et}{\eta}
\renewcommand{\th}{\theta}
\nc{\Th}{\Theta}
\nc{\ka}{\kappa}
\nc{\la}{\lambda}
\nc{\rh}{\rho}
\nc{\si}{\sigma}
\nc{\ta}{\tau}
\nc{\up}{\upsilon}
\nc{\ph}{\phi}
\nc{\ch}{\chi}
\nc{\ps}{\psi}
\nc{\om}{\omega}
\nc{\Ga}{\Gamma}
\nc{\De}{\Delta}
\nc{\La}{\Lambda}
\nc{\Si}{\Sigma}
\nc{\Up}{\Upsilon}
\nc{\Ph}{\Phi}
\nc{\Ps}{\Psi}
\nc{\Om}{\Omega}
\nc{\ptl}{\partial}
\nc{\del}{\nabla}
\nc{\be}{\begin{eqnarray}}
\nc{\ee}{\end{eqnarray}}
\nc{\ov}{\overline}
\nc{\bi}[1]{\bibitem{#1}}
\nc{\fr}[2]{\frac{#1}{#2}}
\nc{\Tr}{\mbox{Tr}}
\nc{\pp}{\partial }
\nc{\ww}{\mbox{\tiny $\wedge$}}
\nc{\sll}{SL(2,${\bf C}$)/SU(2)}
\nc{\ve}{\varepsilon}

\title{A Note on Classical String Dynamics on AdS$_3$} 

\author{M\'aximo Ba\~nados\footnote{max@posta.unizar.es} 
        and Adam Ritz\footnote{aritz@mnhepw.hep.umn.edu}}

\address{
$^*$Departamento de F\'{\i}sica Te\'orica, Universidad de
 Zaragoza, \\ Ciudad Universitaria 50009, Zaragoza, Spain.\\
$^{\dagger}$Theoretical Physics Institute, School of Physics and
Astronomy \\
         University of Minnesota, 116 Church St., Minneapolis, MN
         55455, USA.}
\date{\today}

\maketitle

\begin{abstract}

We consider bosonic strings propagating on Euclidean adS$_3$, and
study in particular the realization of various worldsheet symmetries.
We give a proper definition for the Brown-Henneaux asymptotic target
space symmetry, when acting on the string action, and derive the
Giveon-Kutasov-Seiberg worldsheet contour integral representation
simply by using Noether's theorem. We show that making
identifications in the target space is equivalent to the insertion of
an (exponentiated) graviton vertex operator carrying the corresponding
charge. Finally, we point out an interesting relation
between 3D gravity and the dynamics of the worldsheet on adS$_3$. Both
theories are described by an $SL(2,${\bf C}$)/SU(2)$ 
WZW model, and we prove that the reduction conditions determined 
on one hand by worldsheet
diffeomorphism invariance, and on the other by the Brown-Henneaux
boundary conditions, are the same.  

\end{abstract}

\vfill\eject

\section{Introduction}

Classical three-dimensional gravity was shown by Brown and Henneaux
\cite{BH} to admit an infinite-dimensional conformal symmetry acting
on the space of solutions which are asymptotic to anti-de Sitter space
(adS$_3$). An important feature of this symmetry is that the central
extension, with charge
\be
 c_0 & = & \frac{3l}{2G_3}, \label{c_0}
\ee
where $l$ is the adS$_3$ scale, and $G_3$ is the three-dimensional
Newton constant, arises through the necessity for boundary conditions
at infinity, and not through normal ordering as conventionally occurs
quantum mechanically in two dimensional conformal field theories
(CFTs). The manner in which this symmetry is realized has recently
come under scrutiny after it was pointed out by Strominger
\cite{Strominger97} that a CFT at the boundary, forming a unitary
representation of the Brown-Henneaux symmetry, would have a density of
states sufficiently large to account for the Bekenstein-Hawking
entropy of three-dimensional BTZ black holes \cite{BTZ,BHTZ}. In this
regard, the essentially classical origin of the central charge has
presented an additional challenge for understanding the corresponding
microstates.

More generally, the adS/CFT correspondence \cite{ads,adsrev} implies 
in this instance a duality between string theory on adS$_3$ 
(times some compact space) and a two dimensional CFT. 
In a suitable limit, the correspondence reduces to one between 
supergravity on this space, and a limit of the CFT in which it should 
realize the Brown-Henneaux Virasoro symmetry, with its semi-classical 
central charge. It is then clearly of interest to study string
dynamics
on adS$_3$, in the hope of understanding the microscopic
origin of (\ref{c_0}), and possible quantum corrections to it and the
corresponding black hole entropy. 

One is generally hampered in studying aspects of the adS/CFT
correspondence beyond the supergravity approximation, as the relevant
backgrounds necessarily involve nonzero Ramond-Ramond (RR) sector
fields, for which the worldsheet description is poorly understood
(although see \cite{rr} for recent work). However, adS$_3$ is the
fortunate exception here in that there is a well-defined
10-dimensional string background, corresponding to a configuration of
fundamental strings and wrapped NS5 branes, for which all the RR
fields may be set to zero and a conventional worldsheet description is
possible. Indeed, string theory on adS$_3$ itself has a long history,
initiated in \cite{brfw}, as it provides a useful testing ground for
quantum consistency (see e.g. \cite{su11}). Within the context of the
NS1/NS5 system, and the adS/CFT correspondence, the relevant
background is adS$_3 \times S^3 \times M^4$ (where $M^4$ is either
$T^4$ or K3), and worldsheet aspects of this system were studied
initially by Giveon, Kutasov and Seiberg \cite{GKS}, and have since
been investigated in \cite{BORT,KS,SW,kll,andreev,sugawara}. One
outcome of this work has been the realization, clarified in
\cite{SW,KS}, that string theory on adS$_3$ has, in particular, a
sector of ``long strings''(see also \cite{mms1}) in which the
worldsheet wraps around the boundary of the target space and is
effectively non-compact. This sector is governed by an effective
Liouville system \cite{SW}, and its S-dual formulation in the D1/D5
system was identified in \cite{SW} with the small instanton
singularity in the moduli space of the worldvolume gauge theory. This
latter sector is of particular interest since the worldsheet lives in
the asymptotic region of the target space, and thus should admit
Brown-Henneaux diffeomorphisms as a symmetry. The main purpose of this
note is to study the symmetries associated with worldsheet dynamics on
adS$_3$ and clarify, in particular, the realization of the
Brown-Henneaux target space symmetry. Following the initial work of
\cite{GKS}, this aspect was further elucidated in the work of de Boer
et al. \cite{BORT} who emphasized the distinction between the symmetry
of the first quantized string Hilbert space discussed in \cite{GKS},
and the graviton vertex operators associated with Brown-Henneaux
diffeomorphisms of the target space. In this paper, we shall make the
relation between these two symmetries more explicit in several ways,
and emphasize how the classical origin of (\ref{c_0}) in pure gravity
is mirrored in string theory. 

In particular, we find that the spacetime Virasoro operators of
\cite{GKS} may be obtained directly via a classical Noether
construction associated with the asymptotic symmetry of the worldsheet
action under Brown-Henneaux diffeomorphisms. The charges are conserved
asymptotically in the target space and thus this symmetry applies
rather naturally to the long string sector. We also find that, as in
the gravity description, the central extension (\ref{c_0}) is visible
within classical Poisson brackets. 

A related issue concerns strings propagating on backgrounds which
differ topologically from adS$_3$. Solutions to classical 3D gravity
with negative cosmological constant, or equivalently the low energy
equations of motion of the string, are all locally isomorphic to
adS$_3$, and differ only in their global structure. Starting from
adS$_3$ one may construct these backgrounds via the process of
identifying along discrete symmetries. Since the worldsheet action
inherits the symmetries of the target space, one can make use of an
analogous procedure to obtain strings propagating on more general
backgrounds. Specifically, we shall show that identifying along
discrete symmetries of the worldsheet action changes the topology of
the target space, naturally leading to the identification of the
vertex operators associated with BTZ black holes and conical
singularities.

Several interesting features emerge from the analysis of these
symmetries. In particular, we find that 
the reduction conditions for the 3D Einstein-Hilbert action
associated with the Brown-Henneaux boundary conditions, shown in 
\cite{CHvD} to lead to an asymptotic Liouville dynamics, are actually
equivalent to the constraints imposed by diffeomorphism invariance
on the string worldsheet. As
a consequence, we are able to associate the asymptotic dynamics of
3D gravity, with a constrained worldsheet action for a bosonic string
propagating on $H^+_3=\sll$. 

The plan of the paper is as follows. In Section~\ref{2} we discuss in
some
detail the symmetries of the worldsheet action defined on adS$_3$. 
In particular, we distinguish between the affine worldsheet symmetry
descending from its definition as a coset WZW model, and the
asymptotic target space symmetry associated with Brown-Henneaux
diffeomorphisms. In the latter case, we define an asymptotic Noether
charge in the usual manner, which is shown to be equivalent to the
spacetime generators of \cite{GKS}. In Section~\ref{Vertex} we show
how, on identifying along discrete symmetries, one induces vertex
operators which on exponentiation change the topology of the target
space, leading for example to black hole geometries. In Sec.
\ref{Liouville} we consider the required gauge fixing conditions for
the worldsheet dynamics, and show that the resulting dynamics is that
of Liouville theory, and that in the long string sector the spacetime
generator is directly related to the generator of conformal
transformations on the worldsheet. In Section~\ref{GR-String}, we
explore the relation between the gauge fixed worldsheet dynamics and
the usual reduction of 3D gravity to Liouville theory, via the use of
an auxiliary manifold, whose boundary we identify with the string
worldsheet. We conclude in Section~5 
with some additional remarks on the issue of black hole
microstates.

\section{Classical dynamics of strings on adS$_3$. }
\label{2}

An important property of the Brown-Henneaux conformal symmetry
\cite{BH}, discussed earlier, 
is that the central charge is related, not to normal ordering
ambiguities, but to boundary conditions. For this reason it is visible
classically.
This is also the case in the string theoretic description of this
symmetry, at least in the so-called ``long string" sector in which the
string wraps around the adS$_3$ boundary.  In this
section, we study the classical dynamics of a string propagating on
adS$_3$. We shall give in Sec. \ref{BHS} a proper definition of the
Brown-Henneaux symmetry acting on the string, and derive the contour
integral representation for the generators proposed in \cite{GKS}, as
classical Noether charges.

\subsection{The action and worldsheet symmetries}

The worldsheet theory for bosonic strings propagating on the coset
manifold $H^+_3=\sll$ (or Euclidean adS$_3$) is described by an
$SL(2,${\bf C}$)$ WZW action, 
\be
 I_{W}[h] & = & -\frac{k}{2\pi}\int_{\Si} 
    {\rm tr}(h^{-1}\ptl h)(h^{-1}\ov\ptl h)
     +\frac{ik}{12\pi} \int_{\Si} d^{-1}{\rm tr} (h^{-1}dh)^3,
  \label{WZWaction}
\ee
where $k$ is the WZW level, which throughout we shall take to be
large, 
and the group element
$h$ is restricted to lie in the above coset, the space of $2\times 2$
hermitian complex matrices with unit determinant. This model has two
affine $SL(2,${\bf C}$)$ symmetries, which are identified under
complex conjugation. The generators for these symmetries are the
currents, $J=h^{-1}\ptl h$ and its conjugate $\ov{J}$.

If we parametrise the coset with coordinates $(\ph,\ga,\ov{\ga})$,
\be
 h & = & \left(\begin{array}{cc}
           e^{-\ph}+\ga\ov{\ga}e^{\ph} & e^{\ph}\ga \\
           e^{\ph}\ov{\ga} & e^{\ph}
               \end{array} \right), \label{h}
\ee
then the action for the string worldsheet takes the following form
\begin{equation}
I_W[\phi,\gamma,\bar\gamma] = {k \over 2\pi i}\int_W (\pp \phi
\bar\pp\phi +   e^{2\phi}\pp\bar\gamma\bar\pp\gamma )d^2z.
\label{I}
\end{equation}
For later
convenience, the volume element is taken to be $d^2z=dz\ww d\bar z$,
and
the factor $i$ in the coefficient then ensures that the action is
real.  
Recently, (\ref{I}) has been the object of intense study due to its
important relation with the adS/CFT correspondence
\cite{GKS,BORT,KS,SW,kll,andreev,sugawara}.  
Our aim here is to describe some aspects
of this action which may help to understand its spacetime conformal
properties.

The affine currents above can be decomposed into an $SL(2,${\bf
C}$)$ basis as $J=J^aT_a=kh^{-1}\ptl h/2$, which in the 
parametrization (\ref{h}), and
with a convenient rescaling $\ph \rightarrow \ph/\sqrt{2k}$,
$\ga\rightarrow \ga/\sqrt{k}$ and $\bar\ga\rightarrow
\bar\ga/\sqrt{k}$, take the form,
\begin{eqnarray}
J^- &=& \beta \nonumber\\
J^3 &=&  \beta \gamma + {\alpha \over 2} \pp \phi
\label{currents}\\ 
J^+ &=& \beta \gamma^2 + \alpha \, \gamma \pp\phi + k \pp \gamma,
\nonumber 
\end{eqnarray}
where we have introduced $\beta=-e^{2\ph/\alpha}\ptl\ov{\ga}$ and
$\alpha=\sqrt{2k}$. One recognizes these currents as those arising
from the free field
representation of the affine current algebra \cite{wakimoto}, in which
$\beta$ is treated dynamically. The corresponding action takes the
form, 
\begin{equation} 
I[\phi,\gamma,\bar\gamma,\beta,\bar\beta] = {1 \over 2\pi i} 
\int \left({1 \over 2} \pp \phi \bar\pp\phi - \beta\bar\pp\gamma -
\bar\beta\pp\bar\gamma - e^{-2\phi/\alpha}\beta \bar
\beta\right)d^2z.
\label{Ibeta}
\end{equation}
Varying this action with respect to $\beta$ and $\bar\beta$ one
obtains (\ref{I}) in terms of the rescaled fields. Note that since
we take $k$ large, we work classically and ignore quantum corrections
$\al\rightarrow \sqrt{2(k-2)}$ affecting the screening charge, and 
an induced linear dilaton coupling. For a discussion of
related quantum effects see Refs. \cite{brfw,su11,GKS,KS}.

In the large $\phi$ sector the last term in (\ref{Ibeta}) is a
perturbation and one can use the free OPE representation,
\begin{equation}
\phi(z)\phi(z') \sim -\log(z-z'), \
\ \ \ \ \ 
\beta(z)\gamma(z') \sim {1 \over z-z'}.
\label{OPE}
\end{equation}
Note that the rescaling of $\phi$ is such that its kinetic term
appears divided by two in (\ref{Ibeta}).
This factor is necessary in order to obtain the OPE (\ref{OPE}) which 
then leads to the $SL(2,C)$ algebra at level $k$ for the affine 
currents defined above. These currents are
holomorphically conserved, $\bar\pp J^a=0$, as may be checked 
by using the equations of motion following 
from (\ref{Ibeta}).   

As a consequence of its affine symmetry, the action (\ref{I}) has 
a worldsheet conformal symmetry, corresponding to mappings 
$z\rightarrow f(z)$, with central charge $c=3k/(k-2)\sim 3$. The
stress tensor
which generates the corresponding Virasoro algebra takes the Sugawara
form, $T_W=(J^+J^- - (J^3)^2)/k$, and is (classically)
\begin{equation}
   T_W = \beta\pp\gamma - {1 \over 2} (\pp\phi)^2. 
\label{TW}
\end{equation}
Quantum mechanically, normal ordering generates an improvement
term for $\phi$ which is subleading in $k$ and generates the
central charge $c\sim 3$. 

The dynamics of the string worldsheet is given by the equations of
motion, varying with respect to the fields, plus the
equation,
\begin{equation}
T_W = 0, 
\label{TW=0}
\end{equation}
and its conjugate $\bar T_W=0$, which are required in order to ensure
worldsheet diffeomorphism invariance. Since we focus on classical
aspects of the dynamics we have, in writing this constraint, 
ignored the presence of a product manifold ($M$) in the string
background (adS$_3\times M$), and also additional matter fields. While
these will be required quantum mechanically to ensure vanishing
of the total central charge, they will not affect our
discussion, and we shall now focus on the adS$_3$ sector.  

As we have stressed above, most of the aspects of the Brown-Henneaux
symmetry are visible classically, and thus we could also work at the
level of Poisson brackets. A useful representation for the symplectic 
structure, motivated by the above OPE's, follows by using a
(holomorphic) ``lightcone" decomposition. We shall treat $\bar z$ as
the time coordinate. In this situation, the fields $\bar\beta$ and
$\bar\gamma$ become ``Lagrange multipliers" because they enter in the
action (\ref{Ibeta}) without ``time" derivatives. On the other hand,
the fields $\beta$ and $\gamma$ are canonically conjugate and we
derive the ``equal-time" Poisson bracket, 
\begin{equation}
[\beta(z),\gamma(z')] = \delta(z,z').
\label{P1}
\end{equation}
We define the Dirac delta function as $\oint (dz / 2\pi
i) f(z') \delta(z,z')  = f(z)$.

The field $\phi$ also appears in the action with a linear time
derivative.
In order to extract the Poisson bracket of $\phi$ with itself we need
to assume that the operator $\pp$ is invertible. The invertibility of
$\pp$ is achieved by fixing the value of $\phi$ at infinity $\phi(z
\rightarrow \infty) \rightarrow \phi_\infty$.  An explicit inverse
for this operator can be written as $\pp^{-1} f(z) = \oint (dz/2\pi i)
f(z') \theta(z,z')$ where $\theta(z,z')$ is the step function
satisfying $\pp\theta(z,z')=\delta(z,z')$.  The equal time Poisson
bracket for $\phi(z)$ is then, 
\begin{equation}
[\phi(z),\phi(z')] = - \theta(z,z').
\label{P2}
\end{equation}
The similarity between the OPE's (\ref{OPE}) and the Poisson brackets
(\ref{P1}) and (\ref{P2}) should be evident.  

The Hamiltonian density for this system can be read off directly from
(\ref{Ibeta}). Integrating over a ``spatial slice" yields, 
\begin{equation}
H = \oint {dz \over 2\pi i}\, e^{-2\phi/\alpha} \beta\bar\beta.
\label{Ham}
\end{equation}
(The term $\bar\beta\pp\bar\gamma$ has zero Poisson bracket with all
fields so we omit it.)  It is a straightforward exercise to find the
Hamilton equations, 
\begin{eqnarray}
\bar\pp \beta &=& [\beta(z),H] = 0 \\
\bar\pp \gamma&=& [\gamma(z),H] = -e^{-2\phi/\alpha} \bar\beta\\
\bar\pp \phi  &=& [\phi(z), H] = {2 \over \alpha} \oint {dz'\over
2\pi i} (e^{-2\phi/\alpha} \beta\bar\beta)(z') \theta(z,z'). 
\end{eqnarray}
The first two relations are of course the equations of motion which
follow from varying (\ref{Ibeta}) with respect to $\gamma$ and
$\beta$. The third equation can be put in a more familiar form by
applying
the (invertible) operator $\pp$ to both sides (using
$\pp\theta(z,z')=\delta(z,z')$) obtaining, 
\begin{equation}
\pp\bar\pp \phi ={2 \over \alpha}\, e^{-2\phi/\alpha}
\beta\bar\beta, 
\end{equation}
which is the correct equation of motion for $\phi$.  
The Hamiltonian (\ref{Ham}) plus the canonical relations (\ref{P1})
and (\ref{P2}) represent a (holomorphic) canonical version of the
action (\ref{Ibeta}).  

The conservation of the currents, $\bar\pp J^a=0$, may also
be expressed in Hamiltonian form as, 
\begin{equation}
[J^a(z),H]=0,
\end{equation}
which can be proved directly from the above formulae for $J^a(z)$,
$H$, and the Poisson brackets.  In the same way, one may check
that the currents $J^a$ satisfy the $SL(2,C)$ algebra at level $k$,
and that the worldsheet stress-tensor $T_W$ defined in (\ref{TW})
satisfies the expected classical algebra, 
\begin{equation}
[T_W(z),T_W(z')] = -(T_W(z) + T_W(z')) \pp \delta(z,z'),
\label{TT}
\end{equation}
generating holomorphic worldsheet conformal 
transformations\footnote{Note that this algebra is
equivalent to the OPE, 
$T_W(z)T_W(z')\sim 2T(z')/(z-z')^2+\ptl T(z')/(z-z')$.}.
Quantum mechanically, (\ref{TT}) develops a central charge
$c\sim 3$, as mentioned above.    

Thus far we have focussed on the conventional worldsheet symmetry of
the
action (\ref{I}). However,
since (\ref{I}) describes a string propagating on adS$_3$, one also
expects
a realization of the Brown-Henneaux \cite{BH} spacetime conformal
symmetry, corresponding to mappings $\gamma\rightarrow f(\gamma)$,
with a classical central charge $c \sim 6k$. We shall now analyze this
symmetry in detail.

\subsection{The Brown-Henneaux spacetime conformal symmetry} 
\label{BHS}

In conformal gauge, the action for a string propagating on a curved
background
with metric $G_{\mu\nu}$ and antisymmetric tensor field $B_{\mu\nu}$
takes the usual sigma model form, 
\begin{equation}
I[X] = \frac{k}{4\pi i}
    \int_W d^2 z \left(G_{\mu\nu}(X)\de^{ab}\pp X^\mu_a \bar\pp
X^\nu_b + iB_{\mu\nu}(X)\ep^{ab}\pp X^\mu_a \bar\pp X^\nu_b\right).
\end{equation}
The spacetime symmetries of this action are given by the 
Killing vectors of the spacetime metric $G_{\mu\nu}$ and antisymmetric
tensor $B_{\mu\nu}$\footnote{A Killing vector of $B(X)$ is  
 a transformation of the fields $X$ such that $B$ transforms as
$B\rightarrow B+d\Lambda$ where $\Lambda$ is a 1-form.}. In the
simplest case, the target metric is flat, $G_{\mu\nu}=\eta_{\mu\nu}$,
and the action is Poincar\'e invariant.      

If we now denote $X^\mu=(\phi,\gamma,\bar\gamma)$, and revert
momentarily to the ``geometric field normalization'' of (\ref{I}),
the metric $G$ and field $B$ associated with the action (\ref{I}) are
\begin{eqnarray}
ds^2 &=& d\phi^2 + e^{2\phi}d\gamma d\bar\gamma, \label{adS}\\
B &=& e^{2\phi} d\gamma \ww d\bar\gamma.  
\label{adS/B}
\end{eqnarray}
The metric (\ref{adS}) is a maximally symmetric line element having 6
Killing vectors. These isometries lead to the six dimensional affine
symmetry of the worldsheet action generated by the currents 
(\ref{currents}) and their conjugates.  

As pointed out long ago by Brown and Henneaux \cite{BH}, the line
element (\ref{adS}) has another interesting asymptotic
($\phi\rightarrow \infty$) symmetry given by the transformations, 
\begin{eqnarray}
\delta_{\mbox{{\tiny BH}}} \gamma &=& -\varepsilon(\gamma) \nonumber\\
\delta_{\mbox{{\tiny BH}}} \bar\gamma &=& \frac{1}{2}\,e^{-2\phi} 
 \varepsilon''(\gamma)
\label{BHT}\\
\delta_{\mbox{{\tiny BH}}} \phi &=& \frac{1}{2}\,
\varepsilon'(\gamma) 
\nonumber
\end{eqnarray}
where $\varepsilon(\gamma)$ is an arbitrary function of 
$\gamma$, and
$\varepsilon'$ denotes the derivative with respect to $\gamma$. 
Acting with these transformations on (\ref{adS}) and (\ref{adS/B}) one
generates a new background,
\begin{eqnarray}
 ds^2_{new} & = & d\phi^2+e^{2\phi}d\gamma
 d\bar\gamma+ \frac{1}{2}\varepsilon'''d\gamma^2 \label{dsnew}\\
 B_{new} & = & e^{2\phi}d\gamma \ww
 d\bar\gamma + \varepsilon''d\phi\ww  d\gamma. \label{Bnew}
\end{eqnarray}
The correction for $B$ is clearly a gauge transformation because
$\varepsilon$ depends only on $\gamma$ ($dB_{new}=dB)$. On the other
hand, the correction for the metric is subleading in $\phi$. This is
why the Brown-Henneaux transformations only leave the anti-de Sitter
metric invariant asymptotically.     

Before applying this symmetry to the worldsheet action we pause to
mention an important aspect of the Brown-Henneaux transformations
(\ref{BHT}). Two metrics which differ by a transformation of the form
(\ref{BHT}) are physically distinguishable because the canonical
generator implementing the transformation is a non-zero quantity, the
Virasoro charge \cite{BH}. Since in three dimensions there are no
gravitational waves, it follows that all solutions to the equations of
motion can be generated from anti-de Sitter space via a finite 
Brown-Henneaux transformation.  One can then write the general
solution to the equations of motion with anti-de Sitter boundary
conditions in the form,
\begin{equation}
ds^2 = d\phi^2 +  e^{2\phi} d\gamma d\bar\gamma - {6T(\gamma) \over c}
d\gamma^2 
\label{dsT}
\end{equation}
where $T(\gamma)$ is an arbitrary function of $\gamma$. (One can also
include an arbitrary function of $\bar\gamma$, but for simplicity in
this paper we will restrict all discussions to the holomorphic
sector.) The anti-de Sitter metric corresponds to the vacuum state
with $T=0$ which is $SL(2,C)$ invariant. Acting with the infinitesimal
Brown-Henneaux transformations one generates a non-zero
$T_{new}=-(c/12)\varepsilon'''$, as in (\ref{dsnew}). The line element
(\ref{dsT}) will appear repeatedly in our analysis. See
\cite{Strominger97,Navarro-N,Baires} for recent discussions.   

We now return to the worldsheet action, and consider how the
transformations (\ref{BHT}), which generate asymptotic symmetries of
the background fields (\ref{adS}) and (\ref{adS/B}), are manifested as
symmetries of the action. For this to be the case, it is necessary to
consider strings whose worldsheet is embedded in the region near the
boundary of adS$_3$, i.e. $\phi\rightarrow \infty$. The action
(\ref{I}) is then invariant under the transformations (\ref{BHT}) up
to boundary terms.

In order to make contact with the currents (\ref{currents}) and the
operator found in \cite{GKS}, we shall work with the action
(\ref{Ibeta}) instead of (\ref{I}). The Brown-Henneaux transformations
(\ref{BHT}) are easily translated to the holomorphic sector
$\{\phi,\gamma,\beta\}$. The leading terms are, 
\begin{eqnarray}
\delta_{\mbox{{\tiny BH}}} \gamma &=& -\ve, \nonumber\\ 
\delta_{\mbox{{\tiny BH}}} \beta &=& \ve' \beta + 
          {\alpha \over 2}\, \ve'' \pp\phi, \label{BHT2}\\
\delta_{\mbox{{\tiny BH}}} \phi  &=& {\alpha \over 2}\, \ve'.
\nonumber
\end{eqnarray}
We also note that, under holomorphic transformations,
$\delta_{\mbox{{\tiny BH}}} \ov\beta=0$.  An important property of
these diffeomorphisms is that they leave invariant the worldsheet
stress-tensor, 
\begin{equation}
\delta_{\mbox{{\tiny BH}}} T_W=0,
\label{dT=0}
\end{equation}
where $T_W$ is defined in (\ref{TW}). Since $T_W$ is constrained by
diffeomorphism invariance to be zero, this equation reflects an
important consistency condition. If the product manifold $M$ is added
and $T_W=0$ is replaced by $T_W+T_M=0$, this equation means that the
Brown-Henneaux diffeomorphisms do not mix $T_W$ and $T_M$.

We shall now prove that the action (\ref{Ibeta}) is invariant, up to
boundary terms, under (\ref{BHT2}). For illustration it will be
convenient to take a canonical perspective and view the worldsheet
as interpolating between specified initial and final
closed string configurations. At tree level, a worldsheet without any
additional sources then has a cylindrical topology, and may be mapped
to an annulus within the usual prescription for radial quantization.
In this formulation, one finds that total derivative terms arising
from variations of the worldsheet action reduce to contour integrals
at the initial and final boundaries. Thus the variation of the
action (\ref{Ibeta}) under a Brown-Henneaux diffeomorphism
(\ref{BHT2}) is,
\begin{equation}
\delta_{\mbox{{\tiny BH}}} I = -{\alpha \over 4} 
\sum_j \left(\oint_{C_j} {d\bar z\over 2\pi i}\, \ve' \bar\pp \phi 
+  \oint_{C_j} {dz\over 2\pi i} \, \varepsilon' \pp \phi \right)
\label{deltaI}
\end{equation}
in which the contours ${C_j}$ surround the worldsheet boundaries, and
we have discarded a subleading term in $\phi$. (In the derivation of
(\ref{deltaI}) we have used $\int d^2 z \bar\pp F= \oint dz F$ and
$\int d^2 z  \pp F= -\oint d\bar z F$.)

Via the state--operator map we can always replace the boundary
configurations with operator insertions leading to the conventional
spherical topology with insertions at the boundary points.
However, we shall find that the original canonical picture is more
helpful in the next subsection where we shall consider the Noether
charge for this asymptotic symmetry.

Before turning to this discussion we point out that the above analysis
naturally encompasses the short and long string sectors of the theory.
In the long string sector, where the worldsheet wraps the entire
boundary at conformal infinity ($\ph\rightarrow \infty$) in adS$_3$,
the annulus description for the worldsheet extends to the full complex
plane, and contour integrals surround infinity. In the short string
sector the worldsheet does not wrap the boundary, and may for example
be associated with a correlator of a set of vertex operators inserted
at points $\{z_i\}$ on the worldsheet. As shown in \cite{BORT}, the
worldsheet then develops thin tubes out to conformal infinity,
$\ph\rightarrow\infty$, at each
insertion point. It is clear that in this case, there are additional
boundaries for the worldsheet associated with regularization near each
insertion point. Each insertion point will then lead to an additional
contour integral in (\ref{deltaI}).

\subsection{The Noether charge}
 
Since the variation of (\ref{Ibeta}) under Brown-Henneaux
transformations (\ref{BHT2}) reduces to a boundary term, there is a
Noether
charge associated with this symmetry. As usual, the charge is the
canonical generator of (\ref{BHT2}). In this section we shall 
explicitly compute this charge, which we find to be equal to the 
Giveon-Kutasov-Seiberg operator \cite{GKS}.

Consider a single closed string in an initial state 1, with
coordinates $\{\gamma_1(z),\bar\gamma_1(z),\phi_1(z)\}$, and a final
state 2 with coordinates $\{\gamma_2(z),\bar\gamma_2(z),\phi_2(z)\}$.
(Our arguments will generalize trivially to a more general solution.)
We assume that both $\phi_1$ and $\phi_2$ are large.  
The worldsheet connects the initial and final configurations
generating a cylinder (or some other 
higher genus surface) parametrised by the
``time" coordinate $\bar z$ (see Fig.~1).  As mentioned earlier, a key
assumption we shall need to make is that the initial and final
conditions are such that the full classical worldsheet is contained in
the large $\phi$ region. This assumption will not be fulfilled in
general but it is valid, for example, in the long string sector
\cite{KS,SW}.

Due to the Brown-Henneaux symmetry of the large $\phi$ sector,
there is a conserved charge associated with the above diagram. A quick
derivation of the Noether charge follows by comparing (\ref{deltaI})
with the on-shell variation of the action. 

\begin{figure}
 \centerline{%
   \psfig{file=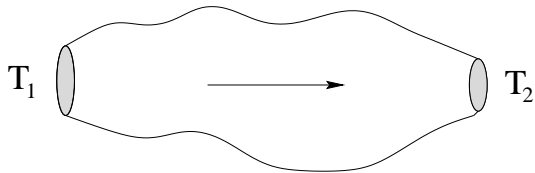,width=7cm,angle=0}%
   }
\vspace{0.1in}
\caption{A schematic representation of a classical (large $\ph$)
worldsheet configuration interpolating two boundary configurations.
Under a Brown-Henneaux target space diffeomorphism, one generates
``charges'' $T_i$ at the boundaries, and the asymptotic symmetry
ensures the conservation condition $T_1=T_2$ within
this diagram.} 
\end{figure}

\noindent

A generic (holomorphic) variation of the action (\ref{Ibeta}) reads, 
\begin{equation}
\delta I = (eom) + {1 \over 2} \sum_j \left( \oint_{C_j} { dz\over
2\pi i} \,  \delta \phi \pp\phi -  \oint_{C_j} { d\bar z\over 2\pi
i}\, \delta \phi \bar\pp\phi -  2 \oint_{C_j} { dz\over 2\pi i} \,
\beta \delta\gamma  \right)
\label{deltaIg}
\end{equation}
where $(eom)$ represents the equations of motion. As before, the
contour integrals are defined at the worldsheet boundaries. In the
situation shown in Fig. 1, the boundaries correspond to the initial
and final configurations and $j=1,2$. The string is closed and for
simplicity we assume that there are no additional insertions, although
as mentioned earlier the initial and final configurations could be
replaced via the state-operator map with two insertions. We now
substitute in this expression the variations (\ref{BHT2}), set the
term $(eom)$ to zero, and compare the result with (\ref{deltaI}). We
find that the contour integrals $\oint d\bar z$ cancel out. The
Noether identity contains only holomorphic contours and reads,
\be
\sum_i T^i_\varepsilon = 0, &&\;\;\;\;\;\;\;
T^i_\varepsilon = -\oint_{i} {dz\over 2\pi i} 
\left({\alpha \over 2}\varepsilon' \pp\phi+\varepsilon\beta \right).
\label{Noether}
\ee 
This relation expresses the on-shell conservation of the ``charge''
$T^i_\ve$.  In the simplest situation with only
two boundaries at the states 1 and 2, the above equation reads
$T^1_\ve=T^2_\ve$ (taking the orientation into account) which
explicitly states the conservation of $T_\ve$.

The conservation equation can also be derived by
the usual trick of letting $\ve$ be a function of $\bar z$, that
is, $\ve=\ve(\gamma,\bar z)$ \footnote{Introducing a dependence on $z$
does not work because the action (\ref{Ibeta}) is actually invariant
under $z-$dependent Brown-Henneaux diffeomorphisms. However, these 
transformations do not define a symmetry of the string theory because
the worldsheet stress tensor $T_W$ is not invariant.}.
Varing the action with this coordinate-dependent parameter, one finds
$\bar \pp t_\ve=0$ with $t_\ve = (\alpha/2) \ve' \pp\phi + \ve \beta$.

As is to be expected, the charge $T_\ve$ generates 
the Brown-Henneaux transformations (\ref{BHT2}) via the Poisson
brackets (\ref{P1}) and (\ref{P2}). Indeed, using (\ref{P1}) and
(\ref{P2}) it is straightforward to check that $\delta_{\mbox{\tiny
BH}}
\gamma(z) = [\gamma(z),T_\ve]$, $\delta_{\mbox{\tiny BH}} \phi(z) =
[\phi(z),T_\ve]$ and $\delta_{\mbox{\tiny BH}} \beta(z) =
[\beta(z),T_\ve]$. The charge $T_\ve$ is also a worldsheet conformal
invariant,
\begin{equation}
[T_W,T_\ve] =0. 
\end{equation}
This relation is just the canonical version of (\ref{dT=0}). 

The charges $T_\ve$ defined in (\ref{Noether}) are infinite in number
because $\ve(\gamma)$ is an arbitrary function of $\gamma$. One would
like to factor out the parameter $\varepsilon$ and extract the charge
as a function of the dynamical variables. However, one notes that the
parameter of the transformation $\ve(\gamma)$ is itself a function of
a dynamical variable.   Let\footnote{A more systematic
approach has been introduced by Teschner \cite{teschner} in which one
organizes representations via a new complex coordinate $(x,\ov{x})$.
This approach has been used in \cite{BORT,KS} and is appropriate for
considering generic vertex operators as functions on adS$_3$. In this
case $\ve(\ga)\rightarrow \ve(\ga-x)$ and one can expand vertex
operators as a powers series in $(x,\ov{x})$ in order to extract the
modes. For our purposes this is unnecessary, although we point out
that all of our results could equivalently be reformulated in this
language.} 
\begin{equation}
\ve(\gamma) = \sum_{n\in N} \ve^n\, \gamma^{n+1},
\label{mode/gamma}
\end{equation}
be a local expansion for $\ve(\gamma)$. Inserting this
expansion into the formula for $T_\ve$ we find,
\begin{equation}
T_\ve = \sum_{n\in N} \ve^n\, L_n, 
\end{equation}
where we have defined, 
\begin{eqnarray}
 L_n &=& -\oint {dz \over 2\pi i} \left({\alpha \over
 2}(n+1)\gamma^n\pp\phi +\gamma^{n+1}\beta \right), \nonumber \\ 
 &=& - \oint {dz \over 2\pi i} \left((n+1)\ga^nJ^3  - n \ga^{n+1}
 J^-\right).
    \label{T/GKS}
\end{eqnarray}
(In the last equality, we have used the formulae for the currents
(\ref{currents}).)  We recognize the operators $L_n$ as the generators
deduced in \cite{GKS}. Note that $L_n$ enters correctly in $T_\ve$ as
the generator of the mode $n$.

The OPE, or commutator of two $L_n$'s can be computed easily using
(\ref{P1}), (\ref{P2}) and yields the Virasoro algebra 
with a central charge \cite{GKS}, 
\begin{equation}
c = 6kp, \ \ \ \ \ \ \ \ 
p := \oint {dz \over 2\pi i} {\pp \gamma \over \gamma}.
\end{equation}   
We stress that this charge arises in both the quantum and classical
calculations.  If the worldsheet $(\ga(z,\ov{z}),\ov{\ga}(z,\ov{z}))$ 
does not wrap around the adS$_3$ boundary, as in the short string
sector, then $p=0$ and thus the central charge is zero, as 
was also pointed out in \cite{BORT}.
Or rather it was argued to arise in this sector from disconnected
worldsheets, therefore requiring a second quantized description. In
this sense, for a single compact worldsheet in the short string
sector, the generators (\ref{T/GKS}) do not correspond directly
to the Brown-Henneaux operators acting on the supergravity action.
We shall make some additional comments on the second quantized
description in Section~V.

The operators $L_n$ first appeared in \cite{GKS}, and were
later discussed in \cite{BORT}, where a derivation of
(\ref{T/GKS}) similar to ours was presented, and also in \cite{KS},
although the full interpretation in terms of conserved charges for the
asymptotic target space symmetry was not given. Within our
construction, since these generators appear as classical Noether
charges, it is clear that they can only be used at large $\ph$,
where they generate a symmetry of the action. Note that this is
a general statement, and not associated with the restricted
validity of the free-field representation for large $\phi$.

\section{Graviton vertex operators and spacetime topology} 
\label{Vertex}

Anti-de Sitter space is the maximally symmetric solution to the string
low energy equations in three dimensions. There exists, however, a
continuum of solutions corresponding to changes in the topology as
well as propagation of Brown-Henneaux ``gravitons" at the boundary.
These solutions are well-known from the gravitational point of view,
and one may consider a string propagating on these backgrounds
\cite{Horowitz-W,Kaloper}. 

In this section we would like to point out that in complete analogy
with the gravitational case, there exists a finite Brown-Henneaux
transformation mapping a string propagating on a black hole to a
string propagating on adS$_3$ with an inserted vertex operator.
This leads to a natural identification of the black hole vertex
operator.

Consider the free string action (\ref{I}), and let $K$ be a Killing
vector of the target adS$_3$ metric. We will identify points along a
discrete subgroup generated by $K$.  From the point of view of the
target metric, this is just the construction of conical singularities
\cite{Deser-J} or black holes \cite{BHTZ} starting from the anti-de
Sitter metric (see \cite{Aminneborg,B3} for generalizations to other
dimensions).  The resulting action will be that of a string
propagating on a conical singularity, or black hole, depending on the
chosen Killing vector $K$.  In the black hole case, it is useful to
write the metric in coordinates such that $K=\pp_\theta$, the
identification then takes the ``natural" form $\theta \sim \theta +
2\pi$, and the metric correspondingly acquires the Schwarzschild form.  

We shall now study the consequences of this construction in string
theory, explicitly working out the simplest case, corresponding
to the generation of an extreme black hole background. In this
section, we consider again the ``geometrical" action (\ref{I}). 
Consider the symmetry of (\ref{I}) under holomorphic dilatations
\begin{equation}
\gamma \rightarrow \lambda \gamma$, \ \ \ \ \ \ \ 
$e^{2\phi} \rightarrow e^{2\phi}/\lambda
\label{sym}
\end{equation}
for a given complex parameter $\lambda$. (Note that $\lambda$ can
depend on $z$, although we consider here the simplest situation in
which $\lambda$ is constant.)  We identify points which differ by the
action of the symmetry,
\begin{equation}
\gamma \sim \lambda \gamma, \ \ \ \ \ \ \lambda \in C.
\label{ident}
\end{equation}
The parameter $\lambda$ is complex and we assume that there exists a
number $p$ such that 
\begin{equation}
 \lambda^p=1.
\label{p}
\end{equation}
If $p$ is real, this means that $\lambda$ is a root of unity, but we
shall not restrict ourselves to this case.  Indeed, for real values of
$p$ the target metric corresponds to a conical singularity, while for
purely imaginary values the identified spacetime has the topology of a
solid torus and corresponds to a black hole. 

Instead of working with the original fields constrained by the above
identifications (as one would do in the orbifold procedure), we make
the field redefinitions,
\begin{eqnarray}
 \gamma'    &=& \gamma^p,  \label{ga'} \nonumber \\
 e^{2\phi'} &=&  { e^{2\phi} \gamma\over p\gamma^p} , \label{newf} \\
 \bar\gamma'&=& \bar\gamma 
      + {1-p \over 2 e^{2\phi}\gamma}, \nonumber
\end{eqnarray}
where $p$ is defined in (\ref{p}). The key property of the new fields
is that the required identifications are now handled automatically
because the combinations $\gamma^p$ and $ e^{2\phi}\gamma$ are
invariant under (\ref{sym}). The action for the new fields takes
the form (we have erased the primes),
\begin{equation}
I_p[\phi,\gamma,\bar\gamma] = I[\phi,\gamma,\bar\gamma] -
\frac{1}{4\pi i} \int d^2z
\, T(\gamma)
\pp\gamma\bar\pp\gamma
\label{iI}
\end{equation}
where $I[\phi,\gamma,\bar\gamma]$ is the free action (\ref{I}) and
$T$ is given by 
\begin{equation}
T(\gamma) = k{1-p^{-2} \over 2\gamma^2},
\end{equation}
which as one might expect is proportional to the Schwarzian 
derivative $\{\ga^{1/p},\ga\}$ associated with the map (\ref{ga'}).
Note that there is also a gauge transformation of the $B$--field,
$\de B = k(p^{-1}-1)/\ga d\ph \ww d\ga$, which we shall, however, ignore
as it corresponds to a total derivative. 

The action (\ref{iI}), resulting from the exponentiation of the
induced vertex operator $V_{\gamma\gamma} = T\pp \gamma
\bar\pp\gamma$, corresponds to a string propagating on the
three-dimensional extremal black hole.  Indeed, the interaction term
in (\ref{iI}) corresponds to adding a piece $T(\gamma) d\gamma^2$ to
the target space metric, as discussed earlier in Eq. (\ref{dsT}).
This is not surprising, as the field redefinitions (\ref{newf}) are a
particular case of a finite Brown-Henneaux transformation (\ref{BHT}).
If we take $p=1+\eta$, with $\eta<<1$, the transformation (\ref{newf})
can be put in the form (\ref{BHT}) with $\varepsilon(\gamma) =
-\eta\gamma \ln \gamma$. Note that in this construction only $L_0$ is
different from zero. Had we taken $\lambda$ as a function of $z$,
which is also a symmetry of the action, we would obtain an action with
other Virasoro charges different from zero.     

One can also introduce the anti-holomorphic transformation $\bar\gamma
\rightarrow \bar \lambda\bar\gamma$ and the solution will no longer be
extremal.  If $p,\bar p$ are real, the identifications induce conical
singularities in the target space. If $p,\bar p$ are purely
imaginary, the target metric is a three-dimensional black hole of
mass $M \sim - (p^{-2}+ \bar p^{-2})$.

\section{Gauge Fixing and Liouville Dynamics}
\label{Liouville}

In section~II, we spent some time considering what happens
within the worldsheet theory when one acts on the target space with
a Brown-Henneaux diffeomorphism. It is now apparent that there are two
symmetries acting on the worldsheet action: the first is the affine
symmetry generated by the $SL(2,${\bf C}$)$ currents $J\times \ov{J}$;
while the second is the conformal symmetry associated with
Brown-Henneaux diffeomorphisms which applies asymptotically in the
target space.

One can ask whether these symmetries are related, and indeed this is
generally true in string worldsheet constructions, in which one
obtains spacetime generators by integrating the corresponding
worldsheet generators against a suitable vertex operator to ensure the
correct conformal dimensions. Within our construction, we have
developed the two symmetries separately, and we would now like to see
whether they are related in this manner. We shall find that this is
indeed the case, once we have imposed the appropriate constraints, and
that the resulting dynamics is that of Liouville theory.

\subsection{Worldsheet Virasoro Algebra}

We begin first with the worldsheet affine algebra generated by 
the currents (\ref{currents}). The additional constraints that must
be imposed are,   
\begin{equation}
kT_W=J^+J^- - (J^3)^2 = 0, \ \ \ \ \ \ \ \  J^-=k,
\label{reds}
\end{equation}
where $J^a$ are the currents (\ref{currents}).  The first equation
expresses diffeomorphism invariance, i.e., the worldsheet
stress-tensor is equal to zero (we ignore here any product manifold
$M$). The second equation is a gauge fixing condition for the residual
conformal symmetry.  

It is interesting to invert the roles of $J^-=k$ and $T_W=0$. Since
$J^-$ commutes with itself, the equation $J^-=k$ can be regarded as a
first class constraint with $T_W=0$ as the ``gauge fixing" condition.
It is well-known that the dynamics of the $SL(2,${\bf C}$)/SU(2)$ WZW
model reduced by $J^-=k$, and its conjugate, yields Liouville theory
\cite{Alekseev-,Forgacs-}.  However, one normally supplements $J^-=k$
with a different gauge fixing condition, namely $J^3=0$. We shall now
prove that imposing $T_W=0$ --leaving $J^3$ arbitrary-- also leads to
Liouville theory, and that the current algebra reduces to a Virasoro
algebra with central charge $c=6k$.

First, as discussed in \cite{Alekseev-,Forgacs-}, it is
straightforward to see that the equations of motion following from
(\ref{Ibeta}), supplemented by the constraint $J^-=k$ and its
conjugate, lead
directly to Liouville theory. Indeed, the field $\phi$ decouples from
$\gamma$ and $\bar\gamma$ and satisfies the Liouville equation,
\begin{equation}
\pp \bar\pp \phi = \al k e^{-2\phi/\al}. \label{liou}
\end{equation}
The role of the ``gauge fixing" constraint is to determine $\gamma$ in
terms of $\phi$.  In our case, we have imposed $T_W=0$ which yields 
\begin{equation}
\pp \gamma = {1 \over 2k} (\pp \phi)^2.
\label{gamma(phi)}
\end{equation}
(We have already replaced $J^-=k$ in this relation.) 

The next step, and perhaps the most important, is to show that under
the reduction conditions (\ref{reds}), the currents (\ref{currents})
lead to a Virasoro algebra with $c=6k$. In the usual case discussed in
the literature, two of the currents, $J^-=k$ and $J^3=0$, are
restricted, while the remaining current $J^+$ satisfies the Virasoro
with
the correct central charge.  In our case, the situation is a bit more
complicated because the constraint $T_W=0$ does not single out a
preferred current to play the role of Virasoro operator.  

The appropriate operator can be found by first twisting the worldsheet
stress tensor $T_W$ in such a way that it commutes with 
the constraint $J^-=k$. As discussed
in \cite{SW} this operator is given by,
\begin{equation}
 Q(z) := T_W - \pp J^3. 
\end{equation}
Imposing the constaints (\ref{reds}), we find 
\begin{equation}
Q(z)= -\pp J^3 = - {1 \over 2} (\pp\phi)^2 - \sqrt{{k \over 2}}\,
\pp^2\phi
\label{J3}
\end{equation}
which we recognize as a stress tensor for the Liouville field
$\phi$, and the coefficient of the improvement term implies that
$-\pp J^3$ generates the Virasoro algebra with $c=6k$ \footnote{One
may wonder what $J^+$ generates after the constaints (\ref{reds}) are
imposed.
$J^+$ is nonzero and, as should be expected for
consistency, generates the same symmetry as $\pp J^3$, although in a
somewhat convoluted manner. From (\ref{reds}), or directly from
(\ref{currents}), we obtain $\pp \sqrt{k J^+} = -Q(z)$ where $Q$ is
given in (\ref{J3}).}.    We shall see in Sec. \ref{GR-String} a
surprising geometrical justification for the appearence of $\pp J^3$.  

Proceeding a little more carefully,
there is actually a technical step in this reduction which should be
checked,
namely, that the Poisson bracket of $\phi(z)$ with itself is still
given by (\ref{P2}), after the gauge is fixed.  This turns out to
be the case because, after inserting the 
constraints (\ref{reds}) into the action, the
kinetic term for $\phi$ does not change. In the Dirac approach, one
constructs the Dirac bracket associated with the second class
constraints (\ref{reds}). Using $[\phi(z),J^-(z')]=0$ one finds
$[\phi(z),\phi(z')]^* = [\phi(z),\phi(z')] = -\theta(z,z')$ as
desired (see the next subsection for more details on the Dirac
bracket). 

It is instructive to consider this reduction at the level of the
action. There is a subtlety here in that the the constraints
(\ref{reds}) imply restrictions on derivatives of the fields, and
care should be taken to include the appropriate boundary terms to
ensure
differentiability of the action. However, since this has been
discussed in \cite{CHvD}, we shall skip the details and write down the
gauge fixed form of the action (\ref{Ibeta}),
\be
 I[\ph]_{gf} & = & \frac{1}{2\pi i} \int d^2 z \left(
    \frac{1}{2}\ptl\ph\ov\ptl\phi - k^2 e^{-2\ph/\al}\right),
\ee  
which we recognize as the Liouville action in conformal gauge, leading
directly to the equation of motion (\ref{liou}). The fields
($\ga,\ov\ga$) are determined through the constraint $T_W=0$ as in
(\ref{gamma(phi)}). Furthermore, if one couples this system to the
background worldsheet metric, necessarily with charge $\al$ for
conformal invariance, one deduces that the stress tensor is given by
(\ref{J3}) as expected.

Thus, after imposing the required gauge constraints, one finds that
the affine worldsheet symmetry is reduced to a local conformal
symmetry with central charge $c=6k$. The dynamics reduces to that of
Liouville theory; in particular we have seen that the stress tensor
has the Liouville form. This discussion fits quite naturally with the
long
string discussion in \cite{SW}, and we note that the reduction is
similar to the conventional reduction to Liouville dynamics of 3D
gravity.  In fact, these two procedures are indeed closely related and
we shall make this connection more precise in Section~IV. However, we
shall now turn to the connection between this Virasoro algebra
associated with the gauge fixed worldsheet dynamics, and the
Brown-Henneaux asymptotic target space symmetry.

\subsection{Worldsheet vs spacetime conformal transformations}

The propagation of strings on adS$_3$ gives rise to two conformal
algebras associated with worldsheet transformations $z\rightarrow
f(z)$, and spacetime transformations $\gamma \rightarrow F(\gamma)$
for large $\phi$. To summarize the worldsheet discussion of the last
subsection, one finds that after the residual conformal freedom is
fixed, and the worldsheet stress tensor is set equal to zero, a
remnant conformal symmetry survives which may be thought of as a
``spectrum generating algebra''. Indeed, the three-dimensional current
algebra $J^a$ is subject to two conditions (see Eq. (\ref{reds})). The
remaining current, which is conveniently chosen to be $Q=-\pp J^3$,
was shown to satisfy the Virasoro algebra with $c=6k$.  Thus, even
after the gauge is fixed we are still left with two Virasoro
operators, the Brown-Henneaux generator $T_{\ve}$ defined in
(\ref{Noether}) (or $L_n$ defined in (\ref{T/GKS})), and the
worldsheet generator $Q$.   

The operators $L_n$ and $Q$ are not in principle related, although
they are both constructed in terms of the currents (\ref{currents}).
In order to compare $L_n$ and $Q$ we need to express $L_n$ in terms of
the gauge fixed variables. Recall, however, that the charges
$L_n$ commute with the worldsheet stress tensor and in that sense
they are diffeomorphism invariant. This means, in particular, that the
algebra of $L_n$ does not change after the gauge is fixed.  

The gauge-fixed version of $L_n$ is obtained by inserting the
conditions (\ref{reds}) into the formula (\ref{T/GKS}) for
$L_n$. This yields, 
\begin{equation}
L_n [\phi] =  -\oint {dz \over 2\pi i} \left( {\alpha \over
2}(n+1)\gamma^n \pp\phi + k \gamma^{n+1}\right).
\end{equation}
We stress here that $\gamma(z)$ should be understood as a functional
of $\phi$ determined by (\ref{gamma(phi)}).  The only dynamical
variable is $\phi$ whose Poisson (or Dirac) bracket is given by
(\ref{P2}).  
Therefore the gauge-fixed $L_n[\phi]$ is a non-local functional (note
that
$T_W=0$ is a differential equation for $\gamma(\phi)$) and generates a
complicated transformation for $\phi$. This should not be a surprise
since it is well-known that working with gauge fixed variables often
yields non-local expressions.

Our task now is to prove that the gauge-fixed $L_n[\phi]$ still
satisfies the Virasoro algebra with central charge $6kp$ in the
Poisson bracket (\ref{P2}).  Below we shall give a rigorous proof of
this fact. There is, however, an alternative way in which
to derive this result which has a nice geometrical interpretation.  

After some direct manipulations using the constraints $T_W=0$
and $\beta=k$, the functional $L_n[\phi]$ can be written 
conveniently as, 
\begin{equation}
L_n[\phi] = \oint {dz\over 2\pi i} \, {\gamma^{n+1} \over \pp\gamma}
\, Q(z) 
\label{LQ1}
\end{equation}
where $Q$ is the worldsheet generator given in (\ref{J3}).  This
formula seems to imply that $L_n[\phi]$ is related to $Q$ simply by
the transformation $z \rightarrow \gamma(z)$. However, this
interpretation is not complete because, on one hand, $\gamma(z)$
is not a function of $z$ but rather a functional of $\phi(z)$, and on
the other, the Schwarzian derivative for the map $\gamma(z)$ is not
present in (\ref{LQ1}). Even though $Q$ is a Virasoro density with
central charge $6k$, the combination (\ref{LQ1}) fails to satisfy the
Virasoro algebra if the dependence on $\phi$, through $\gamma(\phi)$,
is not taken into account.    

Nonetheless, in spite of these subtleties, 
Eq. (\ref{LQ1}) is an identity and we can use it
to relate $L_n$ with the modes of $Q$. Let 
\begin{equation}
Q(z) = \sum {Q_n \over z^{n+2}}
\label{Qn}
\end{equation}
be a local expansion for $Q(z)$. The modes $Q_n$ satisfy the
Virasoro algebra with $c=6k$.  Now consider the sector with 
$p$ long strings, given by $\gamma(z)=z^p+ \rho(z)$. The function
$\rho(z)$ is a small perturbation of the solution $\gamma=z^p$.
Replacing the mode decomposition (\ref{Qn}) in
(\ref{LQ1}) and keeping only the leading term (independent of
$\rho(z)$) we obtain, 
\begin{eqnarray}
L_n \, = \, \sum_m {1 \over p} Q_m \oint {dz \over 2\pi i} \, {1 \over
        z^{1 +  m-pn}} \, = \, {1 \over p} Q_{pn} 
\label{LQ}    
\end{eqnarray}
Since $Q_n$ is a Virasoro operator with central charge $6k$, this
formula implies that $L_n$ has central charge $c=6kp$, in full
agreement with the results of \cite{GKS}.  The formula (\ref{LQ}) 
has appeared previously in the context of
cyclic orbifolds in \cite{Klemm,Borisov}. In \cite{Embeddings,Twisted}
it played a key role in providing a proposal for a statistical
mechanical origin for the three-dimensional black hole entropy. We
shall come back to this issue in the conclusions.  

One point to note is that the transformation (\ref{LQ}) maps
$Q_0,Q_{\pm 1}$ into $L_0,L_{\pm p}$. This implies that the zero mode
of $L_0$ has to be shifted in order to have $L_0,L_{\pm 1}$ as
generators of the $SL(2,\Re)$ subalgebra. This zero mode shift is
equal to the Schwarzian derivative of the map $\gamma=z^p$. The
absence of this shift in the formula (\ref{LQ1}) indicates that the
analysis above is not complete. In particular, we have only obtained
the anticipated algebra with central charge $c=6kp$ for slowly varying
perturbations of the solution $\gamma=z^p$. Below we shall present a
more rigorous discussion and prove that quite generally $L_n[\phi]$
indeed satisfies the Virasoro algebra with the correct central charge
and zero mode.  In fact, the gauge fixed operator $L_n[\phi]$
satisfies the same algebra as the original GKS operator $L_n$. 

This result result follows as a simple consequence of the invariance
of $L_n$ under the action of $T_W$ expressed in the relation
$[L_n,T_W]=0$.  The necessary tool is the Dirac bracket which has the
property that its value does not depend on whether the second class
constraints are imposed or not.  The Dirac bracket associated to the
second class constraints (\ref{reds}) is, 
\begin{eqnarray}
[A,B]^* = [A,B] + {1 \over k}\oint dz \oint dz' [A,\beta(z)]
\theta(z,z') [T_W(z'),B] \nonumber\\ 
+  {1 \over k}\oint dz \oint dz' [A,T_W(z)] \theta(z,z')
[\beta(z'),B]. 
\end{eqnarray} 
and satisfies $[A,T_W(z)]^* = 0 = [A,\beta(z)]^*$ for any function
$A$.  Since $[\beta(z),\phi(z')]=0$ we note, in particular,  
that $[\phi(z),\phi(z')]^*=[\phi(z),\phi(z')]$.  Furthermore,
$L_n[\phi]$ 
only depends on $\phi$ and  thus $[L_n[\phi],L_m[\phi]] =
[L_n[\phi],L_m[\phi]]^* = [L_n,L_m]^*$. This last equality is the
crucial one; the Dirac bracket of the gauge-fixed $L_n[\phi]$ is the
same as the Dirac bracket of the original $L_n$ before the constraints
are imposed. Finally, from $[L_n,T_W]=0$, it is clear that
$[L_n,L_m]^*=[L_n,L_m]$.  The bracket $[L_n,L_m]$ is easy to
compute \cite{GKS} and we find the desired result, 
\begin{equation}
[L_n[\phi],L_m[\phi]] = (n-m) L_{n+m}[\phi] + {c \over 12}
n(n^2-1)\delta_{n+m,0}
\end{equation}
with central charge $c=6kp$.

\section{String dynamics and three-dimensional gravity} 
\label{GR-String}

It was pointed out in Section~IV that the imposition of constraints
such as worldsheet diffeomorphism invariance, naturally reduces the
worldsheet affine symmetry to a Virasoro algebra generated by a
Liouville stress tensor. It was also pointed out that this reduction
has many similarities to the classical Liouville reduction, which
appears in 3D gravity\cite{CHvD}. In this section, we shall explore
this correspondence, and show that the reduction procedures are indeed
equivalent. For this purpose, we shall consider a three-dimensional
manifold $M$ whose boundary, 
\begin{equation}
W = \partial M,
\end{equation}
is to be identified with the string worldsheet (see Fig.~2). 

\begin{figure}
 \centerline{%
   \psfig{file=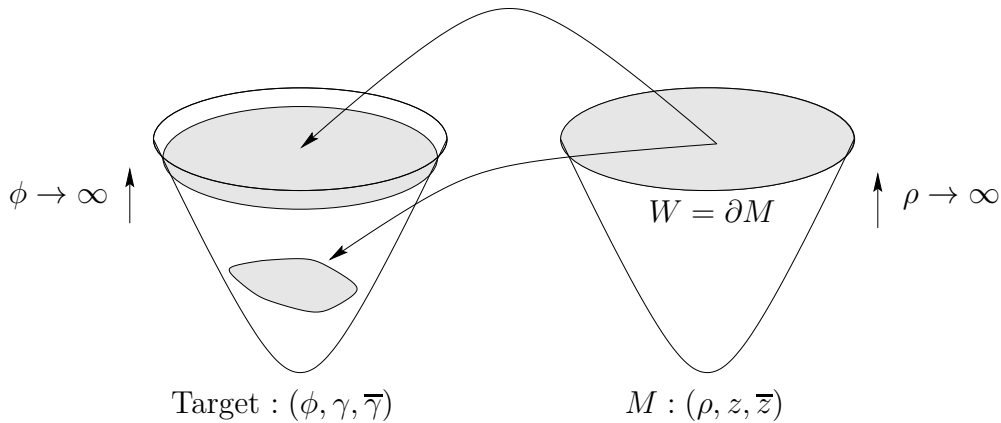,width=10cm,angle=0}%
         }
\vspace{-1in} \hspace{3.7in} $W = \pp M$

\vspace{-0.3in} \hspace{0.35in} $\phi\rightarrow \infty$
               \hspace{4.05in} $\rh\rightarrow \infty$

\vspace{0.9in} \hspace{1.2in} Target $: (\ph,\ga,\ov{\ga})$
               \hspace{1.1in} $M: (\rh,z,\ov{z})$

\vspace{0.1in}

 \caption{A schematic representation of the embedding of the
 worldsheet, $W=\ptl M$ into the target. The upper embedding,
 corresponding to ``long strings'' is the sector which we shall focus
 on here.}
\end{figure}

We will show that a gravitational theory on $M$, with anti-de Sitter
boundary conditions on $W$, is equivalent to the dynamics of the free
worldsheet action (\ref{I}).  Of course, a key step in obtaining this
result is the fact that three-dimensional gravity does not have any
bulk degrees of freedom and all the dynamics lives at the boundary. 

In order to demonstrate this classical equivalence, we first recall
the analysis of \cite{CHvD}, which we shall adapt here to the
Euclidean signature case. Recall that the Euclidean Chern-Simons
formulation of 3d gravity contains two fields $A^a$ and $\bar A^a$
which are complex conjugate. We may introduce the anti-Hermitian
$SL(2,${\bf C}$)$ generators $J_a$ and define $A=A^aJ_a$ and 
$\bar A=\bar A^aJ_a=-A^\dagger$. 
Under the chiral boundary conditions $A_{\bar z}=0$,
$A^\dagger_z=0$, and denoting $z=x^1+ix^0$, the Einstein-Hilbert
action may be written in Hamiltonian Chern-Simons form, 
\begin{equation}
I[A,A^\dagger] = \left[i \int_M (A \ww \dot A + A_0 F) +  \int_W
(A_1)^2\right] +  \left[-i \int_M (A^\dagger \ww \dot A^\dagger
+ A^\dagger_0 F^\dagger) + \int_W (A^\dagger_1)^2\right] 
\label{CS}
\end{equation}
We have suppressed the normalization, and used 
differential form notation in the spatial manifold.
$\dot A$ denotes differentiation with respect to $x^0$, and $F$ is the
field strength. This action has well-defined variations. 
Note that both boundary terms enter the action with the same sign
because the boundary conditions $A_{\bar z}=0$, $A^\dagger_z=0$ imply 
$A_0=iA_1$ and $A^\dagger_0=-iA^\dagger_1$. 

Solving the constraint $F=0=F^\dagger$ on a manifold without 
nontrivial cycles we get $A=g^{-1}dg$,
$A^\dagger=dg^\dagger (g^\dagger)^{-1}$. The action then reduces to
two chiral WZW actions \cite{Moore-S}
\begin{equation}
I[g,g^\dagger] = \left[ iI_{CWZW} (g) + \int (g^{-1}\pp_1 g)^2
\right] + \left[- i I_{CWZW}(g^\dagger)+\int
((g^\dagger)^{-1}\pp_1
g^\dagger)^2  \right].
\label{CWZW}
\end{equation}
We now implement the reduction to a single non-chiral WZW model via
the
argument of \cite{CHvD}, in which the action (\ref{CWZW}) reduces to
\be
 I & = & I_{WZW}[h], \;\;\;\;\;\; h = g^{\dagger}g. \label{WZW}
\ee
More precisely, one may refer to 
\cite{Sonnenschein88} for the details of this reduction in the 
Abelian case\footnote{Consider the action for two fields 
$\phi_L$ and $\phi_R$ with opposite
chiralities, $I[\phi_L,\phi_R] = \int [i\phi'_L \dot \phi_L +
(\phi'_L)^2] + \int [- i\phi'_R \dot \phi_R + (\phi'_R)^2]$. Define
the new variables $\phi=\phi_L + \phi_R$ and $\pi = \phi_L' -
\phi'_R$. The action $I[\phi_L,\phi_R]$ is mapped into $I[\phi,\pi] =
\int [i\dot\phi \pi + {1 \over 2}(\pi^2 + \phi'^2 ) ]$, which is the
Hamiltonian form of a free boson. $\pi$ is eliminated 
using its own (algebraic) equation of motion $i\dot\phi +\pi=0$,
leading to the action for a Euclidean free scalar $I = (1/2)\int[ \dot
\phi^2 + (\phi')^2]$.  The ``$i$'' appearing in the kinetic term is
related to the Euclidean signature. See \cite{Sonnenschein88} for more
details.}, and \cite{CHvD} for the non-Abelian case. 
Note that the equations of
motion for the chiral theories are $\bar\pp g=0$, $\pp g^\dagger=0$
\footnote{Actually, the equations are $\pp_1( \bar \pp g) =0$
but in this situation one normally assumes that $\pp_1$ is
invertible which means that $\pp_1 f=0 \Rightarrow f=0$. This
also eliminates an unwanted gauge symmetry $g\rightarrow ga(t)$ of the
chiral theory.}. The equations for the non-chiral theory are
$\bar\pp(h^{-1} \pp h)=0$ whose general solution is $h=A(\bar z)
B(z)$. If we further impose $h=h^\dagger$ (from (\ref{WZW})) it
follows that $A= B^\dagger$.  

The non-chiral theory has two affine currents
\begin{equation}
J = h^{-1} \pp h, \ \ \ \ \ \  \bar J = \bar\pp h h^{-1}
\label{JJ}
\end{equation}
which are both conserved thanks to the equations of motion. If we
write $h$ in the form $h = g^\dagger g$ with $\bar\pp g=0$ and $\pp
g^\dagger=0$, then $J=g^{-1} \pp g$ and
$\bar J = \bar\pp g^\dagger (g^\dagger)^{-1}$, as expected. This means
that the currents $J$ and $\bar J$ appearing in the non-chiral theory
are indeed the same currents which appear in the chiral theories.

The interesting aspect of this analysis is that since $g$ and
$g^\dagger$ take values in $SL(2,C)$ and the combination $h =
g^\dagger g$ is hermitian, then $h \in SL(2,C)/SU(2)$. The theory is
then precisely the coset WZW model describing a string propagating on
three-dimensional anti-de Sitter space. Using the Iwazawa
decomposition for $SL(2,C)$
\begin{equation}
g = U \left( \begin{array}{cc} e^{-\phi/2} &  0  \\
                               0 &   e^{\phi/2}   \end{array} \right) 
\left( \begin{array}{cc}   1 &  \gamma  \\
                           0 &  1   \end{array} \right)
\label{g}
\end{equation}
where $U\in SU(2)$, $\phi \in \Re$ and $\gamma \in C$, we obtain the
expression (\ref{h}) for $h=g^{\dagger}g$. The non-chiral WZW action
(\ref{WZW}) reduces in this parametrization to the worldsheet
Lagrangian (\ref{I}), thus verifying the following classical relation, 
\begin{equation}
\frac{l}{16\pi G_3}(iI[A^{\dagger}] -iI[A]) = 
    {k_G \over 4\pi i} \int d^2z(\pp\phi\bar\pp\phi + 
    e^{2\phi}\pp\bar\gamma\bar\pp\gamma),
\label{gr-st}
\end{equation}
where the left hand side is the three-dimensional Einstein-Hilbert
action (written in Chern-Simons form), the right hand
side is the string action propagating on Euclidean adS$_3$, and we
have
restored the normalization where $k_G=l/(4G_3)$. The key
step in this identification is that the three-dimensional
gravitational theory is defined on a manifold $M$ whose
two-dimensional boundary $W$ is identified with the string worldsheet. 

Therefore, the $SL(2,C)/SU(2)$ WZW model has two different 
classical interpretations:
(i) as a string theory propagating on adS$_3$; and (ii) as a
gravitational
theory in a three-dimensional manifold whose boundary is identified
with the string worldsheet. Strictly, we are being a little
imprecise in our interpretation since on one hand, a string worldsheet 
requires additional
constraints such as worldsheet diffeomorphism invariance, while on the
gravitational side, one needs to impose the Brown-Henneaux boundary
conditions. Thus, to make this relation complete we need to take these
constraints into account, and a priori it is not at all clear that
they are the same. Nonetheless, we shall now argue that the reduction
conditions imposed by these constraints are indeed equivalent.

We analyzed the reduction of the worldsheet affine currents 
in accordance
with the string constraints in Section~III.A, finding that the
remaining
current could be chosen as $(-\ptl J^3)$ which took the form
(\ref{J3}) and generated a Virasoro algebra with central charge
$c=6k$.
In contrast, as shown in \cite{CHvD}, the gravitational
constraints are,
\begin{equation}
J^3 = 0,  \ \ \ \ \ \ \ \ J^-=k,
\label{redg}
\end{equation}
which ensure that the asymptotic metric satisfies the anti-de Sitter
boundary conditions \cite{BH}.  

Let us study what happens if we impose the string theoretic 
constraints (\ref{reds}) in the gravitational theory.  To
understand the effect of the constraints (\ref{reds}) on the
gravitational variables, it is necessary to recall the general
solution of the gravitational equations containing all currents $J^a$.
This solution can easily be found using the Chern-Simons formulation. 
The general flat connection, solving the Chern-Simons field equations
and satisfying the boundary condition $A_{\bar z}=0$, can be written
in the form (see \cite{Baires} for more details),
\begin{eqnarray}
A_{z} &=& b^{-1} A(z) b, \nonumber\\
A_{\bar z} &=& 0, \\
A_\rh &=& b^{-1} \pp_\rh b \nonumber
\end{eqnarray}
where $b$ depends only on an additional radial coordinate $\rh$.
This solution has an affine symmetry, 
\begin{equation}
A^a(z) \rightarrow A^a(z) + D\lambda^a(z),
\label{affine}
\end{equation}
where $D=\pp +[A(z),~]$ is the holomorphic covariant derivative. In
our parametrization, this symmetry is generated by the
currents (\ref{currents}) which are related to $A(z)$ as follows, 
\begin{equation}
A^a = {J^a \over k}.
\label{AJ}
\end{equation}

Incorporating the other chirality one obtains a solution for the full
Chern-Simons equations, and thus a solution for the gravitational
equations. This solution depends on 6 arbitrary functions, $A^a(z)$
and $\ov{A}^a(\bar z)$, and can be written in the
convenient form, 
\begin{eqnarray}
 ds^2  &=& - A^+A^- dz^2 - \bar A^+ \bar A^- d\bar
z^2 + ( A^- \bar A^+ e^{2\rho} + A^+\bar A^- e^{-2\rho} )
dz d\bar z   \nonumber \\
&& +(d\rho + A^3 dz  + \bar A^3d\bar z )^2.
\label{dsAA}
\end{eqnarray}
It is easy to show that this metric has constant curvature as
required. 
To simplify the notation, we shall assume that the anti-holomorphic
side is
trivial, which means that we impose the conditions $\bar A^+=1,\bar
A^3=0$, and $\bar A^-=0$, in the metric (\ref{dsAA}). With these
simplifications, the metric takes the form,
\begin{eqnarray}
ds^2 = - (A^+A^- - (A^3)^2) dz^2 +  A^- e^{2\rho} dz d\bar z +
d\rho^2 + 2A^3 dzd\rho.
\label{dsA}
\end{eqnarray}
We now proceed to consider the reduction of this metric under the
constraints specified in (\ref{redg}) and (\ref{reds}):

\begin{enumerate}
\item In the conventional reduction in 3D gravity \cite{CHvD}
one imposes the constraints (\ref{redg}) in order to satisfy the 
Brown-Henneaux boundary conditions. The metric then takes the form
(\ref{dsT}) with $T = kA^+= J^+$. On substituting the representation
for
the
currents (\ref{currents}) one finds
\be
 J^+ & = & -\frac{1}{2}(\ptl \ph)^2 -\sqrt{\frac{k}{2}}\ptl^2\ph,
\ee 
as expected for a Liouville stress tensor generating the
Virasoro algebra with central charge\footnote{One may check this
via use, for example, of the free-field commutators for the
currents (\ref{currents}). Alternatively, if one uses the affine
algebra, 
then the reduction is performed via Dirac brackets by
regarding the conditions (\ref{redg}) as a set of second class
constraints. See \cite{Bais-} and \cite{Baires} for explicit
calculations.} $c=6k$.

\item Alternatively, we may also reduce via the constraints
(\ref{reds}), obtaining the metric 
\begin{eqnarray}
ds^2 = e^{2\rho} dz d\bar z + d\rho^2 + 2A^3 dzd\rho.
\label{dsAr}
\end{eqnarray}
This line element is not of the Brown-Henneaux form (\ref{dsT}).
However, one can correct the coordinate $\bar z$ with a term which
is subleading for large $\rh$, 
\begin{equation}
\bar z \rightarrow  \bar z + e^{-2\rho} A^3(z).
\label{bw}
\end{equation}
Replacing this back into the metric (\ref{dsAr}), the cross term
cancels out, and we obtain, 
\begin{eqnarray}
ds^2 = \pp A^3 dz^2 +  e^{2\rho} dz d\bar z + d\rho^2,
\label{dsAr'}
\end{eqnarray}
which has the desired form (\ref{dsT}) with $T =-k \pp A^3 = -\pp
J^3$.  
This is in full consistency with the string reduction (\ref{reds})
where we showed that the combination $-\pp J^3$ played the role of
reduced Virasoro operator with $c=6k$, and in the representation
(\ref{currents}) took the Liouville form (\ref{J3}).
\end{enumerate} 

This analysis indicates that diffeomorphism invariance on the
worldsheet is equivalent to the Brown-Henneaux boundary conditions for
the associated three-dimensional theory. The sets of constraints
(\ref{redg}) and (\ref{reds}) are therefore closely related. One
way in which this relation may be seen is to view the common 
constraint, specifically $\ch=J^--k$, as a first class constraint,
since it commutes with itself. This point was noted earlier in
Section~IV. One may show that if this is the only constraint 
imposed in (\ref{dsA}), one still obtains a correction to the metric,
given by $k((A^3)^2+\ptl A^3-A^+)$, which in terms of the fields
is given by a Liouville stress tensor. This supports the idea that the
constraint $T_W=0$, or alternatively $J^3=0$, acts like a gauge fixing
condition for the symmetry generated by the first class constraint
$\ch=J^--k \approx 0$. 

In interpreting this correspondence, it is natural to assume as we
have
throughout our analysis that $k$
is large since this is the regime of large curvatures and one may use
the low
energy effective action for the string, which in the metric sector
indeed coincides with the 3D Einstein-Hilbert action, plus a
cosmological constant. However, one can also 
think of studying the exact worldsheet theory \cite{GKS,KS}
in the regime where $k$ is small. Since $k\sim 1/\al'$,
this implies that the string effective action should possess large 
$\al'$ corrections, and thus the possibility of a relation to 
pure gravitational dynamics in this regime looks quite nontrivial.

\section{Discussion}

In this paper we have explored several aspects of string dynamics on 
adS$_3$. Our main focus has been on trying to understand the
realization of Brown-Henneaux diffeomorphisms within the 
worldsheet picture, and their relation to affine symmetries of the 
worldsheet action. The connection between these symmetries is 
particularly clear in the long string sector, where the asymptotic 
target space symmetry is realized as a symmetry of the worldsheet 
action. One natural question
to ask is whether this approach sheds any light on the black hole
entropy problem, as it was pointed out by Strominger
\cite{Strominger97}
that a unitary boundary CFT, realizing the Brown-Henneaux symmetry, 
could account for the microstates corresponding to the semi-classical 
Bekenstein-Hawking entropy. We shall conclude in this section, by
making a few remarks on this issue.

If we focus for a moment on the long string sector, then 
it is natural to identify the Brown-Henneaux central charge with that 
obtained within the worldsheet theory \cite{GKS}, with the result,
\begin{equation}
 c_0 = {3l \over 2G_3} = 6kp,
\end{equation}
where we recall that $l$ is the scale of the geometry, and $G_3$ is
the 3D Newton constant. As recalled earlier, the effective
gravitational
dynamics associated with the Brown-Henneaux boundary 
conditions is given by Liouville theory \cite{CHvD} with central
charge $c_0$. We have seen here that the dynamics of the long string 
sector is described by a worldsheet Liouville theory with central
charge $c=c_0/p=6k$. Both theories have an effective level 
density equivalent to a free boson.

It is intriguing to ask what the level density is in a sector
of $p$ coincident long strings. A naive estimate, based on considering
Liouville theory on a multiple cover of the boundary would suggest
a level density associated with p free bosons. This is
also the result we obtain from considering the algebra (\ref{LQ}),
where
we find $Q_0 = pL_0$, which implies that the spacetime generators
$L_n$ form a subalgebra of the $Q_n$'s.  This mechanism has
been explored in \cite{Embeddings} and \cite{Twisted} within pure
three-dimensional gravity.  In the present case, since the
worldsheet central charge is $c=c_0/p=6k$, there is a unitarity
bound on $p$, given by $p\sim c_0$, or somewhat smaller if we 
take into account the
quantisation of $k$ through its interpretation as the 5-brane charge. 
For comparison, we know that a unitary CFT realising the density of
states associated with the BTZ black hole, would also 
have a level density
equivalent to $c_0$ free bosons \cite{Strominger97}.

Of course, this discussion is merely suggestive, as the sector with
$p$ coincident long strings apparently requires a second quantised
desription for a complete analysis. Nevertheless, an interesting
corollary is that in the sector with the highest degeneracy, $k$ is of
order one, (and thus we need to assume $p$ is large in order that
we may ignore quantum corrections -- the string coupling is 
$g^2 \sim 1/p\sqrt{k}$). While this regime is beyond the scope of this
paper, we note that for consistency
the string scale is then of the same order as the adS$_3$ radius,
since
$k=(l/l_s)^2\sim 1$ implies $l \sim l_s$. This is appropriate for the 
long string sector if all fluctuations are on the scale of the
characteristic adS$_3$ size, implying that short strings are somehow
not excited. 

If we are in a regime where all fluctuations are at the adS$_3$
scale $l$, then one can ask how one should distinguish between 
unwound (short) and wound (long)
strings. In recent work \cite{sugawara}, it has been suggested that
this distinction may be associated with the
presence or otherwise of an abelian flux on the string worldsheet,
via the description of the NS1/NS5 configuration within matrix string
theory. It was also suggested there that the multiply-wrapped case in
the long string sector ($p>1$) should indeed be interpreted as $p$
wrapped
strings. As discussed above, this 
naturally implies the necessity for a second quantized
framework,
or matrix string theory. Some generalization of this or a related kind
seems necessary in order to describe the short and long string sectors
in a unified manner.

Another issue, closely related to the speculations
above, is whether the spectrum arising from the dynamics of
long strings is in fact related to the spectrum of black holes. 
One hint in this direction
is that Liouville theory is known to possess a threshold in the
spectrum beyond which it is continuous. This threshold can be deduced
directly from the coefficient of the improvement term in the stress
tensor (e.g. (\ref{J3})).
One finds \cite{SW},
\begin{equation}
 \De \sim  \frac{k}{4} \sim  \frac{c}{24},
\end{equation}
which is the Ramond-Ramond
ground state, associated with zero-mass extreme black holes. Thus
it would be nice to associate the continuum above this threshold
with the spectrum of black holes. For some work on the identification
of black hole states in the effective boundary Liouville theory
of pure gravity see \cite{martinec}. Examination of the 
elliptic genus \cite{deboer} (see also \cite{mms})
also suggests that the ability of supergravity to match the spectrum
via the adS/CFT correspondence breaks down at this threshold, implying
the
need for stringy corrections.

\subsection*{Acknowledgments}

MB thanks M. Asorey and F. Falceto for many useful conversations, and
we also thank D. Kutasov for helpful comments and R. Argurio for
pointing out a small error in the original preprint. MB
was supported by CICYT (Spain) grant AEN-97-1680, and the Spanish
postdoctoral program of Ministerio de Educaci\'on y Cultura, and the
work of AR was supported in part by the Department of Energy under
Grant No. DE-FG02-94ER40823.

\end{document}